\documentclass[conference]{IEEEtran}
\usepackage{times}

\usepackage[numbers]{natbib}
\usepackage{multicol}
\usepackage{hyperref}

\usepackage{graphicx}
\usepackage{multirow,array}
\usepackage{subfig}
\usepackage{amsmath} 
\usepackage{amssymb}  
\usepackage{svg}

\usepackage{booktabs}

\usepackage{amsthm}
\newtheorem*{definition}{Definition}

\author{\IEEEauthorblockN{Jack Geary}
\IEEEauthorblockA{School of Informatics\\
University of Edinburgh\\
Email: Jack.Geary@ed.ac.uk\\}
\and
\IEEEauthorblockN{Henry Gouk}
\IEEEauthorblockA{School of Informatics\\
University of Edinburgh\\
Email: henry.gouk@ed.ac.uk}}

\title{Altruistic Decision-Making for Autonomous Driving with Sparse Rewards}
\date{June 2020}

\begin{document}

\maketitle

\begin{abstract}
In order to drive effectively, a driver must be aware of how they can expect other vehicles' behaviour to be affected by their decisions, and also how they are expected to behave by other drivers. One common family of methods for addressing this problem of interaction are those based on Game Theory. Such approaches often make assumptions about leaders and followers in an interaction which can result in conflicts arising when vehicles do not agree on the hierarchy, resulting in sub-optimal behaviour. In this work we define a measurement for the incidence of conflicts, Area of Conflict (AoC), for a given interactive decision-making model. Furthermore, we propose a novel decision-making method that reduces this value compared to an existing approach for incorporating altruistic behaviour. We verify our theoretical analysis empirically using a simulated lane-change scenario.
\end{abstract}

\IEEEpeerreviewmaketitle

\section{Introduction \& Related Work}
\begin{figure}[t]
    \centering
    \subfloat[\label{intro_figure_image}]{%
        \includegraphics[height=0.25\textheight]{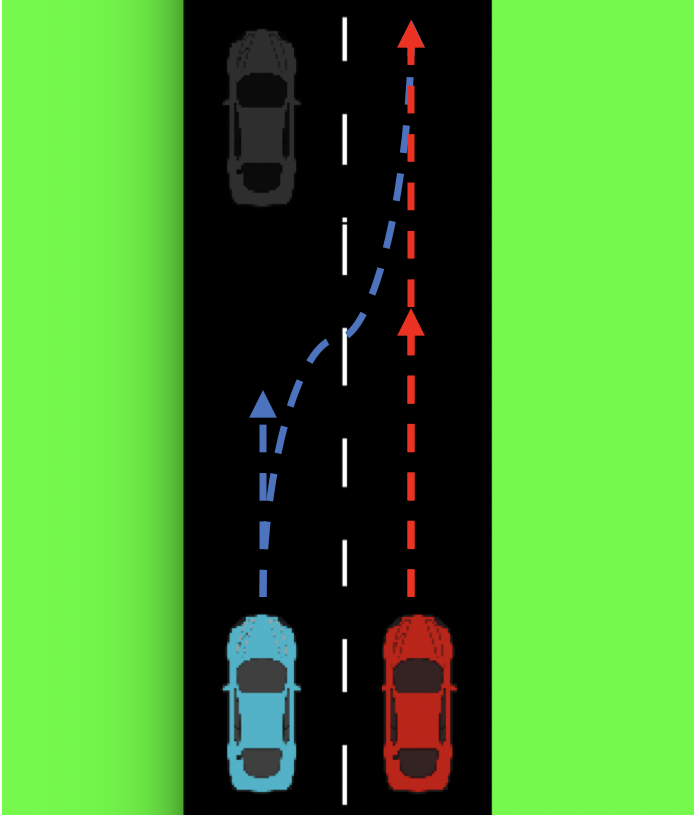}}
    \newline
    \begin{flushleft}
        \subfloat[\label{intro_figure_table}]{%
            \begin{tabular}{cc|c|c|}
                & \multicolumn{1}{c}{} & \multicolumn{2}{c}{$C2$}\\
                & \multicolumn{1}{c}{} & \multicolumn{1}{c}{$GW$}  & \multicolumn{1}{c}{$C$} \\\cline{3-4}
                \multirow{2}*{$C1$}  & $D$ & $(-\infty,-\infty)$ & $(0,1)$ \\\cline{3-4}
                & $LC$ & $(1,0)$ & $(-\infty,-\infty)$ \\\cline{3-4}
            \end{tabular}}
    \end{flushleft}
\caption{(a): Motivating Example; Car C1 (blue) is in the left lane approaching a stopped car (grey). Car C2 (red) is in the adjacent lane. Dotted lines depict the options available to each vehicle; C1 cam either change lanes (LC) ahead of C2, or decelerate (D) to avoid collision. C2 can either continue at current velocity (C) or give way (GW) to allow C1 to merge onto the lane. (b) Reward matrix associated with the motivating example; C2 would prefer to continue, and C1 would prefer to change lanes. Diagonal entries in table represent states where either both vehicles collide or neither agent's objective is satisfied, which is mutually undesirable.}
\label{intro_figure}
\end{figure}

Driving is an interactive task that requires agents to make decisions about if and when to perform certain manoeuvres in the pursuit of a navigational goal. Completing these manoeuvres often requires coordination with other drivers on the road without direct communication. Approaches to this in the autonomous driving literature typically presume to have access to a dense reward that has features that account for the interaction between agents \citep{Sadigh2018, fisac2019, Schwarting2019}. These reward functions are generally  learnt or hand-crafted before deployment---a process that requires manually specifying features that might be relevant. The reliance on complicated reward functions is a limitation of existing techniques that could pose serious safety risks or verification issues.

Game Theoretic approaches are a broad class of techniques that can be utilised to perform decision-making in problems involving interactions between agents. While many methods exist, in autonomous driving literature, it is common to treat such problems as Stackelberg games \citep{sadigh2016, Sadigh2018, fisac2019, Schwarting2019}. This provides a computationally tractable method for computing a joint equilibrium strategy for all the interacting agents. However, these approaches rely on a known hierarchy identifying the leader and followers in an interaction (e.g. at an intersection where vehicles enter in the order of arrival, at a signalised junction) which, in practise, is unknown. In general it is not the case that such a hierarchy is known (e.g. during a lane merge where drivers must simultaneously agree to execute a manoeuvre). This creates a problem: agents using the same decision-making assumptions can arrive at conflicting conclusions about the optimal strategy due to disagreements about who the leader and followers are. Such conflicts can result in sub-optimal or unsafe behaviour. We propose a metric---the Area of Conflict (AoC)---for quantifying the extent to which a pair of decision-making models will result in conflicting assumptions about who takes the role of leader or follower. This metric can be seen as a measure of robustness to the level of aggressiveness or passiveness exhibited by other agents in the driving environment.

Using the insight provided by our conflict analysis, we propose a method for interactive decision-making that requires less familiarity with the other vehicles on the road than previous approaches, while also being less susceptible to conflicts. Our approach makes use of a sparse reward signal, defined using the intended goals of each road user. By including terms in the reward function that correspond to the success of other agents in accomplishing their goals, we are able to effectively model interactive behaviour between different agents on the road. This concept of altruism originates in Game Theory literature. \citet{andreoni1993} presents the idea of altruism being a scalar value, $\alpha$, that multiplies or adds with the rewards of the interacting agents to influence an agent's decision-making by the potential payoffs to the other agents. In the work, \citeauthor{andreoni1993} provide three distinct models for altruism: pure, duty and reciprocal. Our proposed definition best aligns with the definition for pure altruism. Similar to altruism in Game Theory, there is Social Value Orientation (SVO) in the fields of psychology and behavioural economics \citep{mcclintock1989}. \citet{Schwarting2019} implement a version of SVO that is similar to our proposed altruism implementation in that they weight the planning agent's reward, and the rewards of the other agents according to the planning agent's SVO value.

\section{Conflict in the Stackelberg Game}
In a 2-person Stackelberg game one player takes the role of the leader and the other the role of the follower. The leader chooses the action that maximises their reward under the assumption that the follower will behave optimally with respect to the leader's choice. For example, using the reward matrix given in Figure \ref{intro_figure_table}, if $C1$ were the leader then they would choose to lane change (and get reward of 1) as, when the follower, $C2$, chooses, aware $C1$'s choice, they will choose to give way (and get a reward of 0). However, if $C2$ were the leader instead, they would choose to continue (and get reward 1) and $C1$ would be forced to decelerate. Thus a conflict can emerge if it has not been agreed in advance which agent is the leader and which agent is the follower. In the case of autonomous driving, without any means of direct communication, or even an agreed upon policy, no such agreement can be reached. We define conflict as follows.

\begin{definition}
    Conflict: When players of a Stackelberg game compute different equilibria due to uncertainty over the identity of the leader and the follower.
\end{definition}

Conflicts can be problematic as they can result in unforeseen catastrophic situations. In our example, if both $C1$ and $C2$ decide on an action under the assumption that they are the leader, the resulting situation will be ($LC$,$C$), and the vehicles would crash. Therefore it is important that the method for decision-making has as low an incidence of conflict possible, so that it is equipped to handle a diverse range of other agents. In our example it is clear from the reward matrix that the players are in conflict, and there is no clear way to resolve it.\

\section{Altruism and Area of Conflict}
\label{altruism_and_area_of_conflict}
In this section we will define our variant of altruism, as well as an augmentation to the definition to account for iterated planning. We will also provide a definition for Area of Conflict.\ 

\subsection{Altruism}
\begin{figure}[t]
    \centering
    \begin{tabular}{cc|c|c|}
        & \multicolumn{1}{c}{} & \multicolumn{2}{c}{$C$}\\
        & \multicolumn{1}{c}{} & \multicolumn{1}{c}{$B1$}  & \multicolumn{1}{c}{$B2$} \\\cline{3-4}
        \multirow{2}*{$R$}  & $A1$ & $(r_{111},r_{112})$ & $(r_{121},r_{122})$ \\\cline{3-4}
        & $A2$ & $(r_{211},r_{212})$ & $(r_{221},r_{222})$ \\\cline{3-4}
    \end{tabular}
\caption{General reward matrix}
\label{reward_matrix}
\end{figure}

We model the interactive driving problem as a simultaneous game played on a reward matrix, indexed by actions, where each cell in the matrix contains the rewards received by each player if they each chose the corresponding action.  Figure \ref{reward_matrix} presents a general reward matrix where if the row player, $R$, and column player $C$, chose actions $A1$,$B1$ respectively, $R$ would receive a reward of $r_{111}$ and $C$ would receive a reward of $r_{112}$. The grid is $2 \times 2$ for demonstrative purposes and, in general, the grid can be of any size $M \times N$ where $M$ is the number of actions available to $R$ and $N$ is the number of actions available to $C$, and each cell contains a reward pair $(r_{mn1},r_{mn2})$. For simplicity we focus on the case where there are only 2 players however, this can generalise to any number of players. Unless the full index is required, we will refer to the row player's reward for a particular action combination as $r_{1}$ and, correspondingly, $r_{2}$ for the column player's reward.\

Pure Altruism, as defined in \cite{andreoni1993}, makes use of an altruism coefficient $\alpha$ to define the altruistic reward,
\begin{equation}
    r^{*}_{i} = r_{i} + \alpha r_{-i} \quad 0 \leq \alpha \leq 1,
\end{equation}
where the $-i$ index corresponds to the agent that is not indexed by $i$, and $r^{*}_{i}$ is the effective reward agent $i$ uses to perform decision-making. If $\alpha=0$ then the agents are indifferent to one another and if the value is $1$ then the agents are cooperating in order to maximise the same reward, $r^{*}_{i} = r_{i} + r_{-i}$. 

As an alternative to Pure Altruism we propose the following definition for the altruistic reward:
\begin{equation}
    r^{*}_{i} = (1-\alpha_{i})r_{i} + \alpha_{i}r_{-i} \quad 0 \leq \alpha_{i} \leq 1.
    \label{eqn:altruism_definition}
\end{equation}
In this case each agent has their own individual altruism coefficient, and scaling agent $i$'s reward in parallel with agent $-i$'s allows for more flexible behaviours; if $\alpha_{i}=0$ then the agent is wholly egoistic, if $\alpha_{i}=1$ then the agent is wholly altruistic. To avoid confusion we will refer to the \cite{andreoni1993} altruism as ``pure altruism" and our proposed definition as ``altruism".\ 

Extensive previous and ongoing work has been dedicated to estimating reward functions and interactive parameters \citep{albrecht2016,albrecht2019,albrecht2020,Schwarting2019}. In this work we presume that the ``true" reward matrix $\{(r_{mn1},r_{mn2})\}_{0 < m \leq M, 0 < n \leq N}$, and altruism values $\alpha_1,\alpha_2$ are known to both agents. Each agent can then, independently, construct the reward matrix $\{(r{*}_{mn1},r^{*}_{mn2})\}_{0 < m \leq M, 0 < n \leq N}$, which they will use to choose which action to perform. 

\subsection{Augmented Altruism}

Repeated iteration over the system of equations defined by Equation \ref{eqn:altruism_definition} produces a variation that accounts for both agent's awareness of the other altruistic coefficient. By finding the steady-state of this system we arrive at the definition of the altruistic reward presented in Equation \ref{eqn:augmented_altruism_definition}. In the interest of saving space we defer the complete derivation to Appendix~\ref{deriving_augmented_altruism}.
\begin{equation}
    r_{i}^{*} = \dfrac{(1-\alpha_{i})r_{i}+\alpha_{i}(1-\alpha_{-i})r_{-i}}{1-\alpha_{i}\alpha_{-i}} \quad i \in \{1,2\}
    \label{eqn:augmented_altruism_definition}
\end{equation}

This value, which we refer to as the augmented altruistic reward, is an improvement on our base altruism definition, as it is a computationally tractable method for accouting for both players altruism values when evaluating options, whereas the original definition only accounted for the agent's own $\alpha$.

\subsection{Area of Conflict}
Altruism can be used to resolve conflict scenarios; in the example in Figure \ref{intro_figure_table}, if $\alpha_{1}=1$, for instance, then $C1$ would get an effective reward of $0$ for performing the lane change, and a reward of $1$ for decelerating and allowing $C2$ to proceed. However, altruism does not entirely eliminate conflict since, if $\alpha_{2}=1$ also, then the players are once again in conflict, with each so eager to facilitate the other that neither can achieve their objectives.\ 

Let $f_{I}:\mathbb{R}^{M \times N} \times [0,1] \times [0,1] \rightarrow \{A1,A2\} \times \{B1,B2\}$ be a function, parametrised by the altruism coefficients, mapping from the reward matrix to the equilibrium of the corresponding Stackelberg game for some interactive decision-making model $I$. The previous observation indicates that, for a given reward matrix, there is a region in the parameter-space  that will always result in conflict. We call this region the Area of Conflict (AoC). It is desirable to choose a decision-making method that minimises the AoC for a given reward matrix.\ 

In the following derivations we will refer to the reward matrix defined in Figure \ref{reward_matrix}. Without loss of generality we will assume the cell $(A2,B1)$ is optimal for $R$ and $(A1,B2)$ is optimal for $C$. We further assume that there are no ambiguities in each agents' rewards. This gives us the following:

\begin{equation}
    \begin{split}
        r_{211}>&r_{121},r_{111},r_{221} \\
        r_{122}>&r_{212},r_{112},r_{222}.
    \end{split}
\end{equation}
It is immediately clear that with these constraints that decision-making on the reward matrix will result in conflict, regardless of the value of the parameters. Therefore it is vacuously true that the AoC for the traditional Stackelberg solution method is $1$ \cite{von2010}. We will use this value as a baseline. \ 

In general we observe that conflict will occur if:
\begin{equation}
    \begin{split}
        & (r^{*}_{211}>r^{*}_{121} \land r^{*}_{122}>r^{*}_{212}) \\
        \lor & (r_{211}^{*}<r_{121}^{*} \land r_{122}^{*}<r_{212}^{*})
    \end{split}
    \label{eqn:conflict_definition}
\end{equation}

The definition of the AoC of a decision-making model follows from Equation \ref{eqn:conflict_definition}, and we defer the explicit derivations to Appendix~ \ref{deriving_area_of_conflict}. The AoC definitions for the standard Stackelberg Game, Pure Altruism, SVO, Altruism, and Augmented Altruism are provided in Table \ref{AoC_general_table}. Table \ref{AoC_specific_table} presents evaluations corresponding to the reward matrix in Figure \ref{intro_figure_table}. 

\begin{table}[t]
    \centering
    \caption{AoC definitions for various interactive decision-making models, where we set $A = r_{211}-r_{121}$ and $B = r_{122}-r_{212}$ for compactness. See Appendix~\ref{deriving_area_of_conflict} for the definitions of $p_1$ and $p_2$.}
    \label{AoC_general_table}
    \begin{tabular}{ll}
        \toprule
        & Area of Conflict\\
        \midrule
        Baseline \cite{von2010} & $1$   \\
        Pure Altruism \cite{andreoni1993} & $min(\frac{A}{B},\frac{B}{A})$\\
        SVO \cite{Schwarting2019} & $\frac{p_{1}p_{2} + (\frac{\pi}{2}-p_{1})(\frac{\pi}{2}-p_{2})}{(\frac{\pi}{2})^{2}}$   \\
        Altruism & $2(\frac{AB}{(A+B)^{2}})$   \\
        Aug-Altruism & $ln(A+B)(\frac{A}{B} + \frac{B}{A}) - (\frac{A}{B}ln(A) + \frac{B}{A}ln(B)) - 1$ \\
        \bottomrule
    \end{tabular}
\end{table}

\begin{table}[t]
    \centering
    \caption{Calculated AoC values for various interactive decision-making models based on the reward matrix given in Figure \ref{intro_figure_table}.}
    \label{AoC_specific_table}
    \begin{tabular}{lc}
        \toprule
        & Area of Conflict\\
        \midrule
        Baseline \cite{von2010} & $1$   \\
        Pure Altruism \cite{andreoni1993} & $1$\\
        SVO \cite{Schwarting2019} & $0.5$ \\
        Altruism & $0.5$   \\
        Aug-Altruism & $.38623$ \\
        \bottomrule
    \end{tabular}
\end{table}

\begin{figure}[t]
    \begin{center}
    \subfloat[\label{a_vs_aoc_b_1}]{
        \includegraphics[width=.3\textwidth]{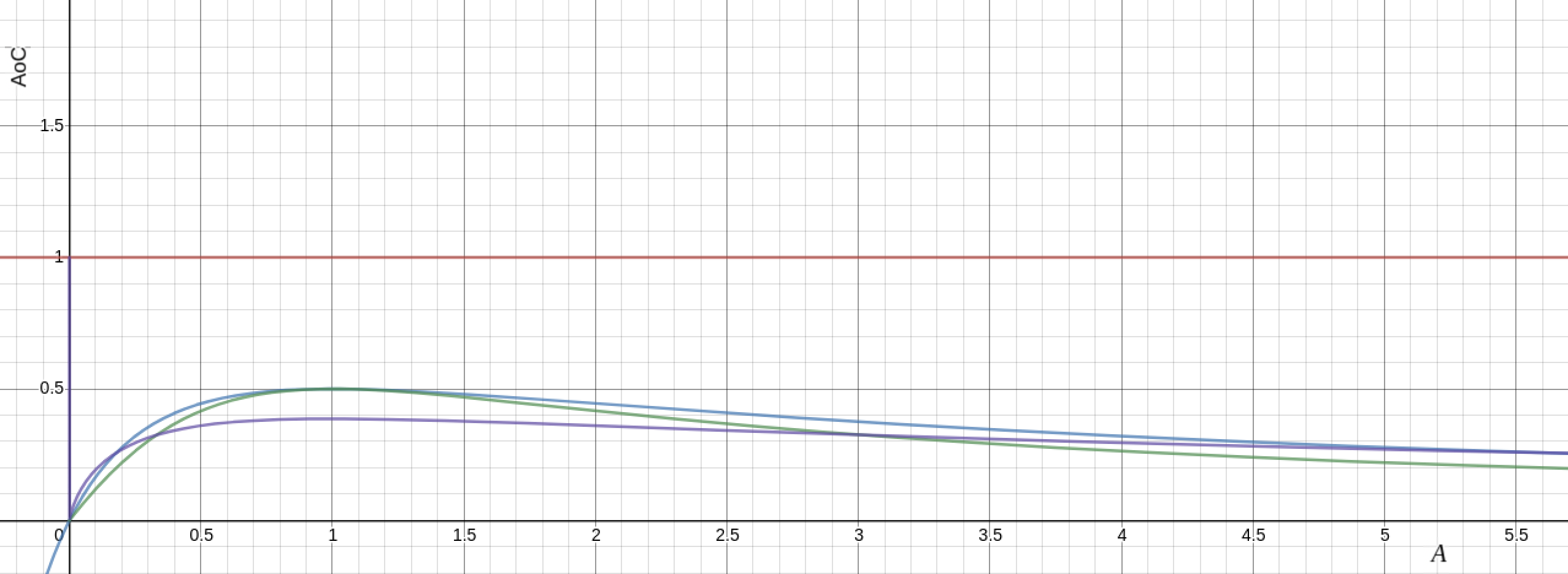}}
    \end{center}
    \begin{center}
    \subfloat[\label{a_vs_aoc_b_3_5}]{
        \includegraphics[width=.3\textwidth]{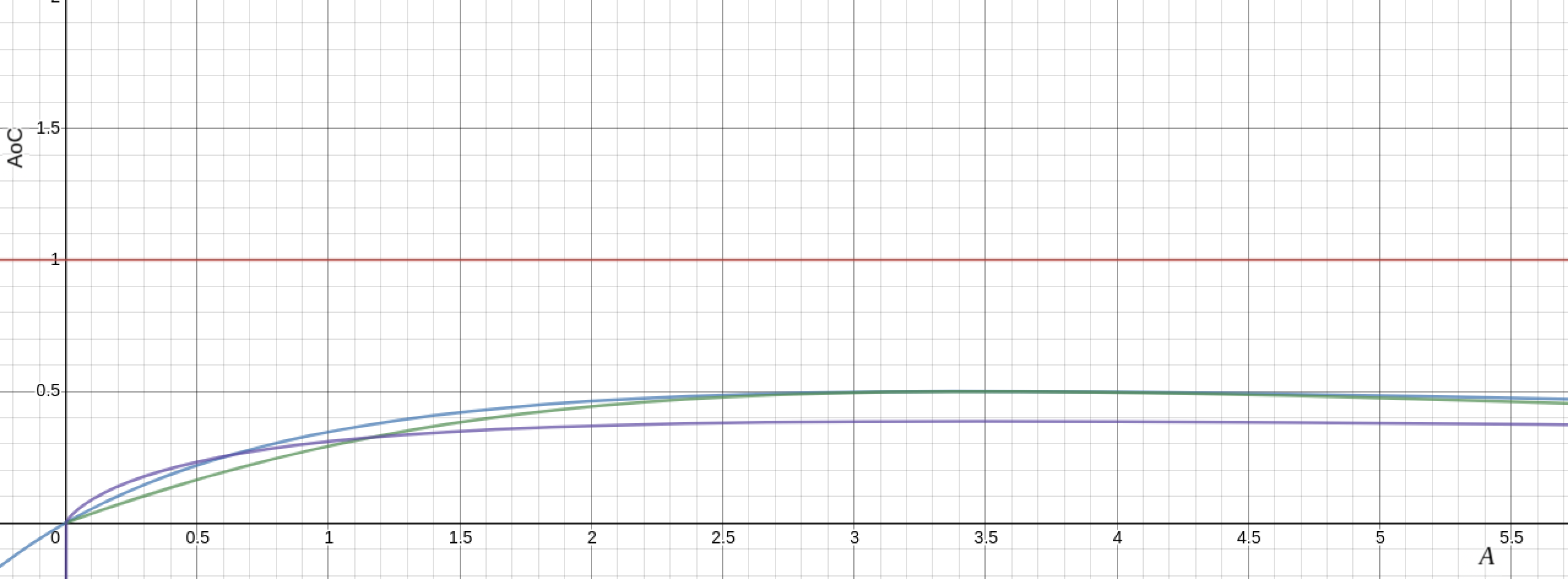}}
    \end{center}
    \caption{
    Plot of the AoC for varying values of $A$. The Blue line corresponds to Altruism, the Green SVO, and the Purple Augmented Altruism. (a) $B=1$, (b) $B=3.5$.  }
    \label{altruism_aoc_plots}
\end{figure}

We observe that the AoC for the Augmented Altruism significantly outperforms the other considered models. This means that, in repeated pairings of agents with altruism scores sampled uniformly from $[0,1]$, the incidence of conflict would be lowest when using this model. In general we empirically observe that, for reasonable magnitudes of $\frac{A}{B}$ Augmented Altruism consistently outperforms the other models. Figure \ref{a_vs_aoc_b_1} shows the AoC plotted against $A$, when $B=1$. We observe that for $0.33<A<3$ Augmented Altruism achieves minimal values. From Figure \ref{a_vs_aoc_b_3_5} we see that when $B=3.5$ this range is $1.6<A<10.4$. This demonstrates the effectiveness of the proposed model for minimising conflict. In the next section we will demonstrate how this result influences the ability for optimal control planners to plan and execute optimal trajectories.
\section{Experiments}

\begin{figure}[h]
    
    \subfloat[\label{res_Alt}]{\includegraphics[width=.25\textwidth]{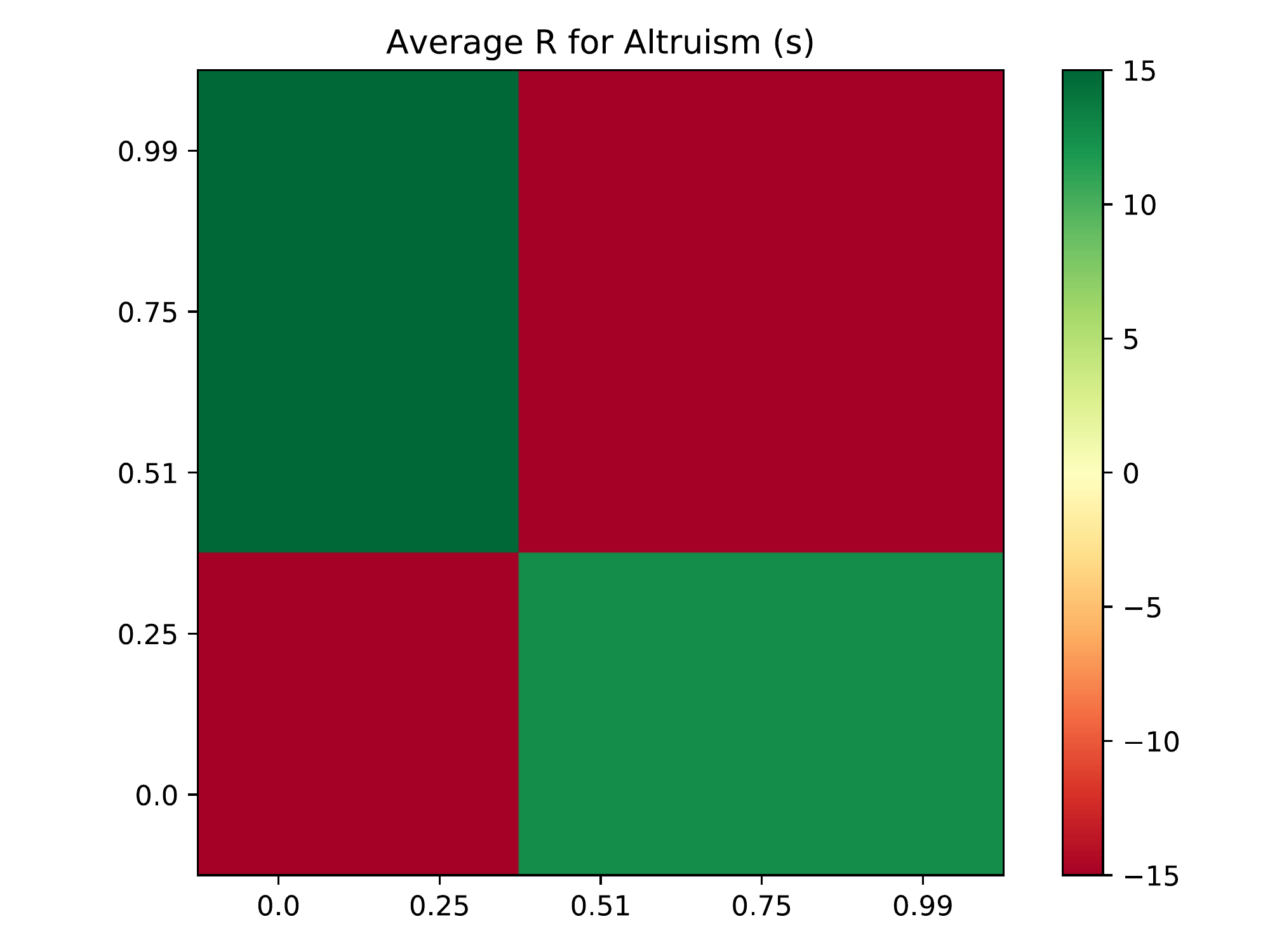}}
    \subfloat[\label{res_SVO}]{\includegraphics[width=.25\textwidth]{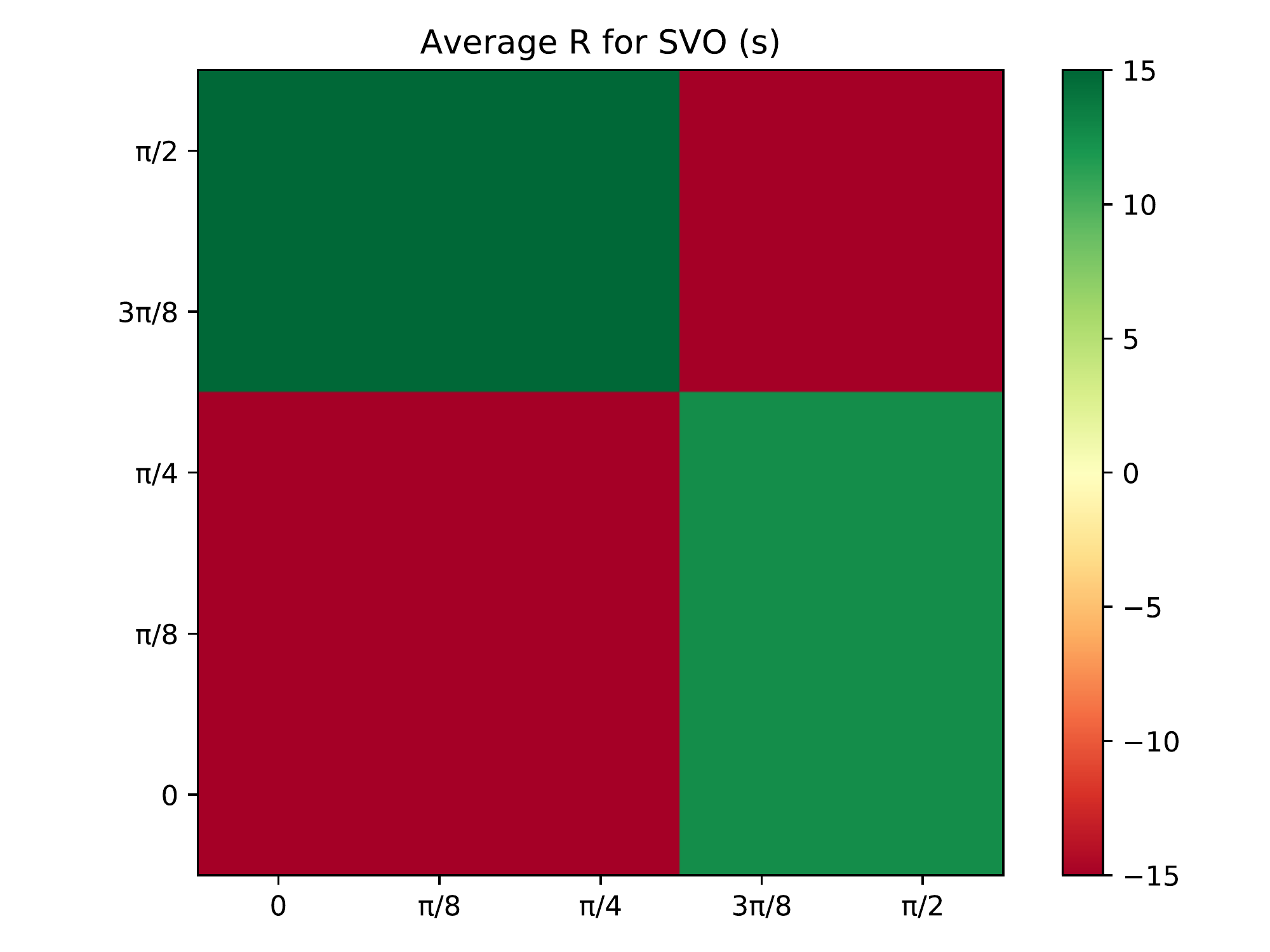}}
    \\
    \subfloat[\label{res_AugAlt}]{\includegraphics[width=.25\textwidth]{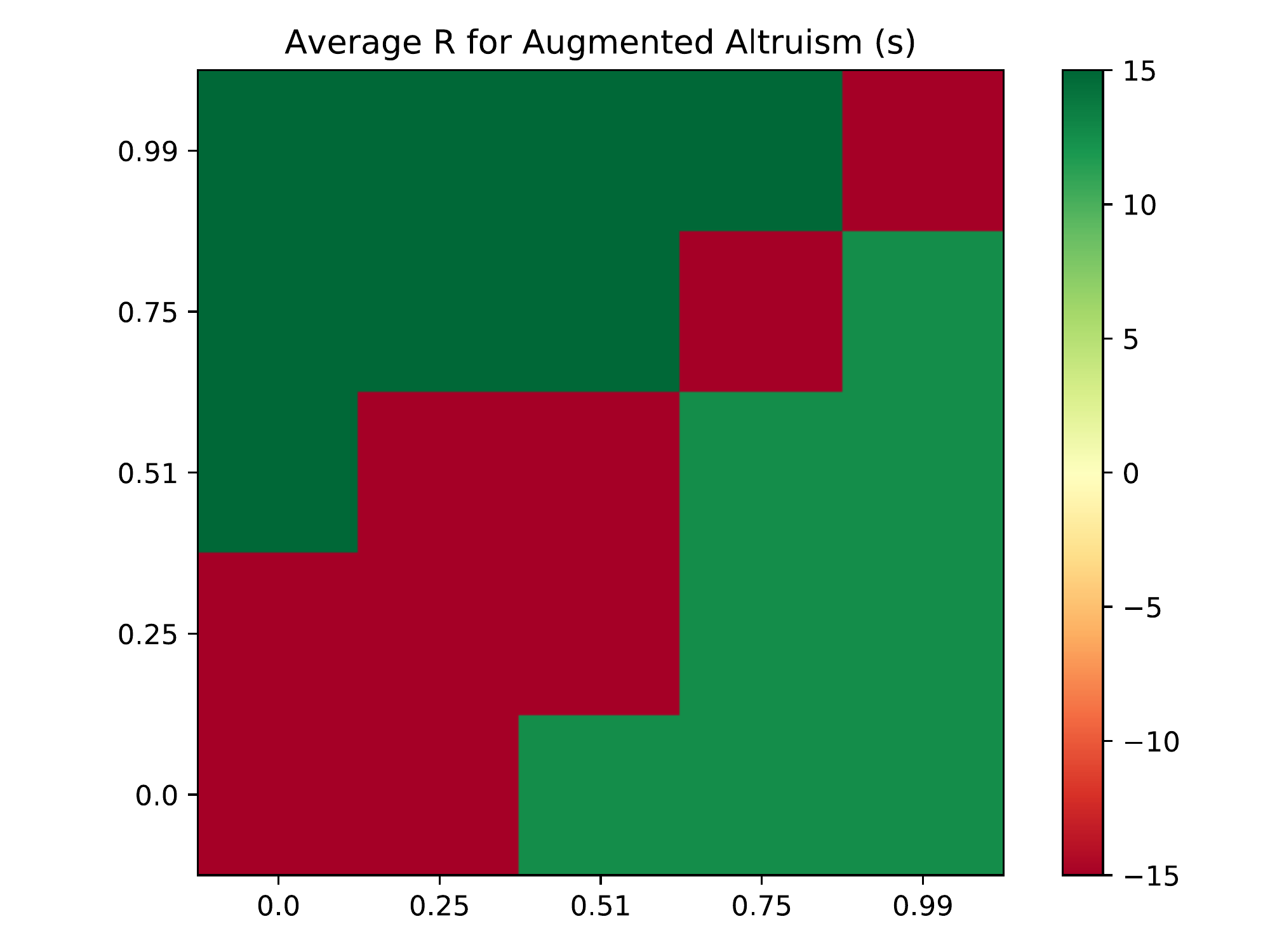}}
    \subfloat[\label{res_base}]{\includegraphics[width=.25\textwidth]{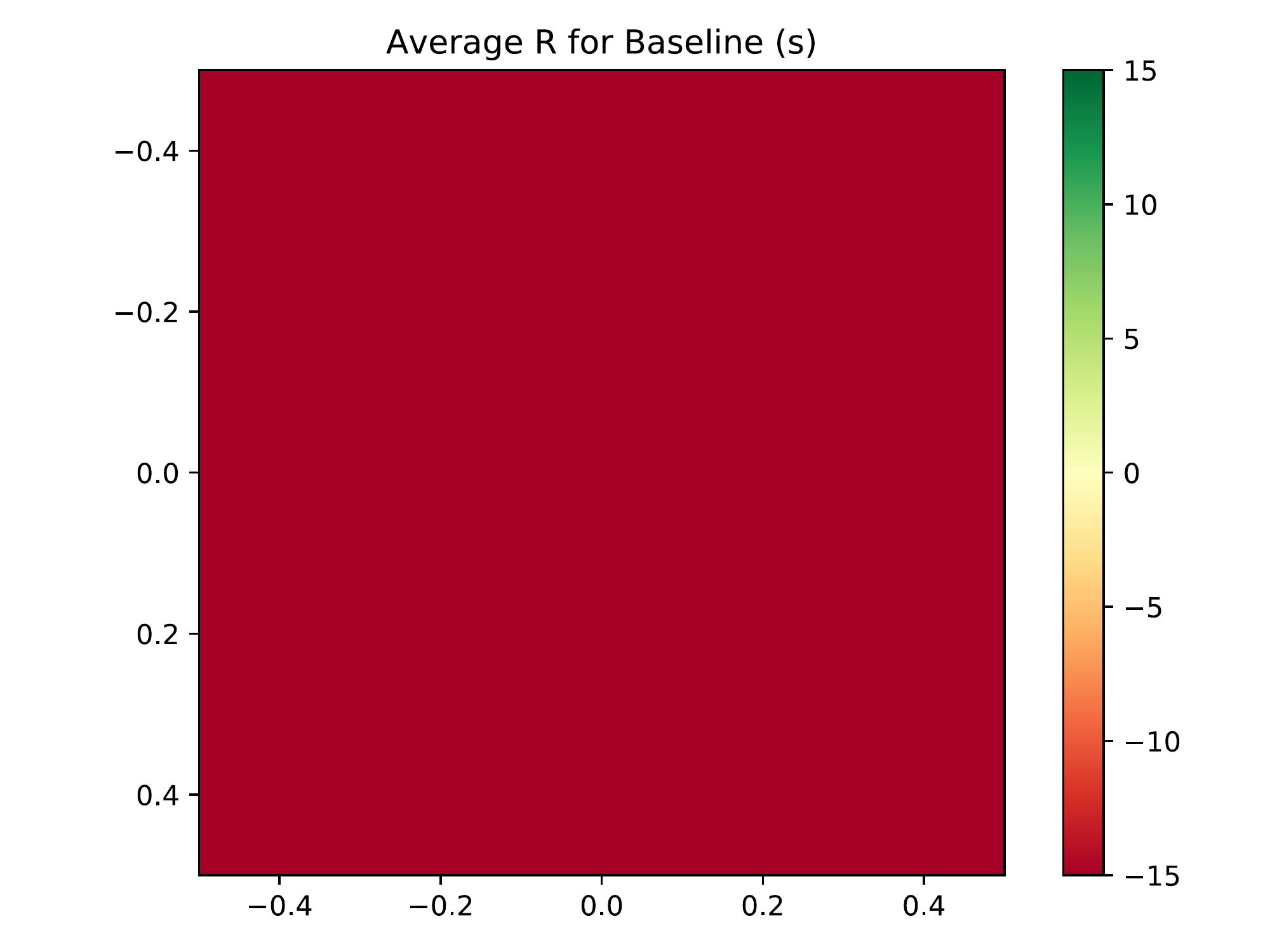}}
    \caption{Plot of the average total signed time required for the vehicles to reach their objective. The trajectory time is signed using the maximum reward received by either player during the experiment. Each $(\alpha_{1},\alpha_{2})$-indexed cell in the grid contains the average result achieved over 25 experiments. The order of the results are; (a) Altruism, (b) SVO, (c) Augmented Altruism, (d) Baseline }
    \label{results_plots}
\end{figure}

The theoretical analysis provided in Section~\ref{altruism_and_area_of_conflict} demonstrates the extent to which each method is robust to conflicts. In this section we show how conflicts can impact the ability of an agent to accomplish its objective in a timely fashion. The experimental setup is described in Appendix~\ref{experimental_method}, and the specific dynamics model used during simulation is given in Appendix~\ref{planning}.

The measurements taken during the experiments were the time taken for each vehicle to get within a distance $\epsilon$ of their intended destination state, $t_{c}^{i}$, and the true reward pair $(r_{1},r_{2})$ received by the players based on the executed actions $(A_{i},B_{j})$. This value is computed from the ground truth reward matrix. To report the results we sum the completion times, and multiply this by the maximum reward received by either player for the execution,
\begin{equation*}
    \begin{split}
        T_{c} &= t_{c}^{1} + t_{c}^{2} \\
        R &= max(r_{1},r_{2}) \times T_{c},
    \end{split}
\end{equation*}
which encodes both the time taken to reach the goal state (magnitude) and whether a conflict occurred (sign).

From the construction of the experiment we would expect that conflict will result in both agents being too passive, or both being too aggressive. In the former, both will choose the less desirable action to allow the other to proceed, and in the latter both agents will pursue their objective, expecting the other to give way. We would expect, when the players are in agreement, that trajectory durations should be relatively small, as each agent has a good approximation for the other agent's trajectory.\ 

The results for our experiments are shown in Figure \ref{results_plots}. The x and y axes of the grid are parameterised by $C1$ and $C2$'s interactive parameter values respectively. The colour of each cell in the grid is determined by the average value of $R$ achieved over repeated experiments.\

The trends of the grid roughly approximate the shapes cut out by the constraint boundaries depicted in Figure \ref{constraint_bound_plot}, which provides an empirical verification of our theoretical analysis. As predicted by the AoC figures reported in Table~\ref{AoC_specific_table}, the baseline method that does not perform any form of altruistic modelling is always in conflict. In contrast, the three altruistic planners are able to produce conflict-free plans with varying degrees of success. The SVO and Altruism methods obtain comparable number of conflicts (13 cells with conflicts, 12 without). Furthermore, our Augmented Altruism method is able to achieve the lowest number of conflicts, with nine conflict cells and 16 non-conflict cells. This shows that our method is more robust than other approaches to the underlying levels of altruism that can be exhibited by agents.
\section{Conclusion}
In this work we proposed a novel formulation for an interactive decision-making model based on existing theories in the Game Theory literature. We proposed a method for augmenting such models, and a novel metric, AoC, for measuring and comparing the performance of an interactive decision-making model. Using this method we demonstrated that our augmented model achieved better theoretical and empirical results than existing methods. We also demonstrated that conflict negatively impacted a planning agent's ability to construct efficient, interactive, trajectories, and that a lower AoC correlated with shorter trajectories. 

\section*{Acknowledgements}
The authors would like to acknowledge Subramanian Ramamoorthy for facilitating this work. Funding for this work was provided in part by FiveAI.

\bibliographystyle{plainnat}

\bibliography{references}

\clearpage
\begin{appendices}
    \section{Deriving Augmented Altruism}
\label{deriving_augmented_altruism}
When attempting to identify equilibria in Game Theoretic problem formulations it is not uncommon to used iterative best response methods to compute the Nash Equilibrium \cite{vorobeychik2008}. In practise this involves each agent choosing an optimal action based on the optimal actions for the other agents in the previous iteration. It is known that if these methods converge to a solution, it is a Nash Equilibrium \cite{bacsar1998}.\ 

We observe that in the context of applying altruism, an iterative approach can be used; after each agent computes their altruistic reward once, they can repeat the process using the rewards computed in the previous iteration. This gives us the following system of equations:

\begin{equation*}
    \begin{split}
        r_{1}^{k}  &= (1-\alpha_{1})r_{1} + \alpha_{1}r^{k-1}_{2} \\
        r_{2}^{k}  &= (1-\alpha_{2})r_{2} + \alpha_{2}r^{k-1}_{1},
    \end{split}
\end{equation*}
where $k\geq 0$ gives the iteration index, and we assume that agent $i$ does not iterate over reward $r_{i}$ as the amount of reward they would get from achieving their own objective, $(1-\alpha_{i})r_{i}$, is not a value that needs to be optimised. Since the altruism coefficients are bounded, $0 \leq \alpha_{i} \leq 1$, we know this system will converge (we assume that $\alpha_{1}$ and $\alpha_{2}$ are not both exactly $1$, as this renders the computation unsolvable). We can find the steady state for this system by solving:

\begin{equation*}
    \begin{split}
        r_{1}^{\infty} &= (1-\alpha_{1})r_{1} + \alpha_{1}r^{\infty}_{2} \\
        r_{2}^{\infty}  &= (1-\alpha_{2})r_{2} + \alpha_{2}r^{\infty}_{1} \\
    \end{split}
\end{equation*}

Which yields the solution:

\begin{equation*}
    r_{i}^{*} = \dfrac{(1-\alpha_{i})r_{i}+\alpha_{i}(1-\alpha_{-i})r_{-i}}{1-\alpha_{i}\alpha_{-i}} \quad i \in \{1,2\}
\end{equation*}

\section{Deriving Area of Conflict}
\label{deriving_area_of_conflict}
Conflict will occur if:
\begin{equation}
    \begin{split}
        & (r^{*}_{211}>r^{*}_{121} \land r^{*}_{122}>r^{*}_{212}) \\
        \lor & (r_{211}^{*}<r_{121}^{*} \land r_{122}^{*}<r_{212}^{*})
    \end{split}
    \label{eqn:conflict_definition_supp}
\end{equation}
By solving these inequalities for the various methods listed above we get the following bounds for the AoC (In order to save space we will let $A = r_{211}-r_{121}$ and $B = r_{122}-r_{212}$).:

\begin{itemize}
    \item{Pure Altruism}: $(\alpha_{1}>\frac{A}{B} \land \alpha_{2}>\frac{B}{A}) \lor (\alpha_{1}<\frac{A}{B} \land \alpha_{2}<\frac{B}{A})$
    \item{Altruism}: $(\alpha_{1}>\frac{A}{B+A} \land \alpha_{2}>\frac{B}{B+A}) \lor (\alpha_{1}<\frac{A}{B+A} \land \alpha_{2}<\frac{B}{B+A})$
    \item{Augmented Altruism}: \\
            ($1-\frac{1-\alpha_{1}}{\alpha_{1}}\frac{A}{B}<\alpha_{2}<\frac{B}{B+(1-\alpha_{1}A)}$ $\land$  $0<\alpha_{1}<1$)
\end{itemize}

\begin{figure}
    \centering
    \includegraphics[width=\columnwidth]{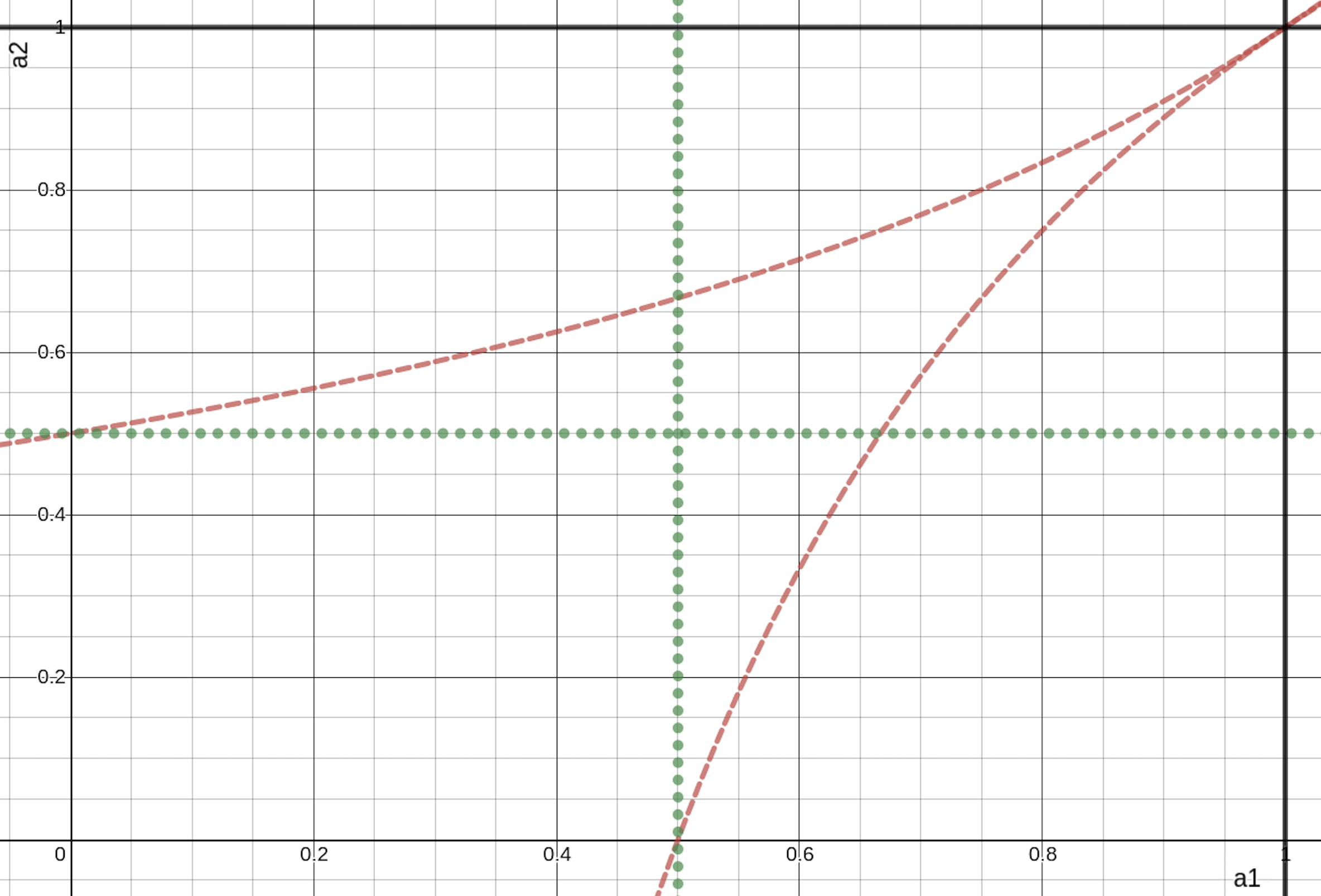}
    \caption{A plot of the constraints defining AoC for various reward models for the example reward matrix (Figure \ref{intro_figure_table}). Per the Altruism constraints (Green) the game is in conflict if the coefficients (a1,a2) are in the antidiagonal quadrants. The same bounds apply for SVO when $\theta_{i}$ is normalised. For the Augmented Altruism constraints (Red) the game is in conflict if (a1,a2) lie in the region bounded by the x and y-axes and the red dotted lines.}
    \label{constraint_bound_plot}
\end{figure}

In all of the above cases it also holds that $0 \leq \alpha_{i} < 1$, except in the case of augmented altruism, where $0<\alpha_{1}<1$. Each of the logical conjunctions ($\land$) specifies a bounded region of parameter space which will result in conflict, and the logical disjunctions ($\lor$) define pairs of non-overlapping regions (see Figure \ref{constraint_bound_plot} for a graphical depiction of these regions). Therefore we can define the AoC as the sum of the areas of these regions in parameter space. For Pure Altruism, and our proposed Altruism, these are straightforward computations.The AoC for Augmented Altruism is given by:

\begin{equation}
    \begin{split}
        AoC & = \int_{0}^{1}\frac{B}{B+(1-\alpha)A}d\alpha - \int_{\frac{A}{A+B}}^{1} 1-\frac{1-\alpha}{\alpha}\frac{A}{B} d\alpha \\
        &= ln(A+B)(\frac{A}{B}+ \frac{B}{A})- (\frac{A}{B}ln(A) + \frac{B}{A}ln(B)) - 1
    \end{split}
\end{equation}\ 

For comparison we can also perform the same evaluation for SVO (\cite{Schwarting2019}).

\begin{equation}
    r_{i}^{*} = \cos(\theta_{i})r_{i} + \sin(\theta_{i})r_{-i} \quad 0\leq \theta_{i} \leq 2\pi 
\end{equation}

By the same procedure as before we observe that conflict occurs with SVO when:
\begin{equation}
    \begin{split}
    &(\theta_{1}<\tan^{-1}(\frac{A}{B}) \land \theta_{2}<\tan^{-1}(\frac{B}{A})) \\ 
    \lor & (\theta_{1}>\tan^{-1}(\frac{A}{B}) \land \theta_{2}>\tan^{-1}(\frac{B}{A}))
    \end{split}
\end{equation}

Even though the SVO mechanism allows for masochistic and sadistic behaviours (corresponding to angles resulting in coefficients with negative magnitudes), to facilitate comparison we constrain the SVO coefficients to be between 0 and 1. This  implies $0<\theta_{i}<\frac{\pi}{2}$. We can therefore compute the AoC for SVO as:

\begin{equation}
    \begin{split}
        p_{1} &= max(0,min(\frac{\pi}{2},\tan^{-1}(\frac{A}{B}))) \\ 
        p_{2} &= max(0,min(\frac{\pi}{2},\tan^{-1}(\frac{B}{A}))) \\
        AoC &= \frac{p_{1}p_{2} + (\frac{\pi}{2}-p_{1})(\frac{\pi}{2}-p_{2})}{(\frac{\pi}{2})^{2}}
    \end{split}
\end{equation}

\section{Trajectory Planning}
\label{planning}
In the literature the problem of interaction-aware trajectory planning for autonomous vehicles is often treated as an optimal control problem, under the assumption that an accurate, dense reward function is available for the other interacting vehicles (\cite{sadigh2016},\cite{fisac2019},\cite{Schwarting2019}). In practise this is generally an overly conservative assumption, as such reward functions often require a familiarity with the subject that is typically not available. In order to demonstrate the efficacy of our proposed decision-making method we articulate the path planning problem as an optimal control problem with a generic reward, that uses the information from the interaction-aware decision-making models to coarsely estimate the behaviour of the other agent.\ 

\subsection{Vehicle Model}
We model the dynamics for both vehicles using a discrete kinematic bicycle model. This is given by:

\begin{equation*}
    \begin{bmatrix}
        x_{k+1}\\y_{k+1}\\v_{k+1}\\\theta_{k+1}
    \end{bmatrix}
     = 
    \begin{bmatrix}
        x_{k}\\y_{k}\\v_{k}\\\theta_{k}
    \end{bmatrix}
    +
    \begin{bmatrix}
        v_{k}cos(\theta+\delta_{k})\\v_{k}sin(\theta_{k}+\delta_{k})\\a_{k}\\\frac{2v_{k}}{L} sin(\delta_{k})
    \end{bmatrix}
    \Delta t
\end{equation*}

where $L$ is the inter-axle length of the vehicle, and $(a_{k},\delta_{k}) \in \mathbb{R}^{2}$ are control inputs received from the planner. 

\subsection{Modelling Other Vehicles}
To model the behaviour of the other interacting vehicle we assume that each action available to the vehicle corresponds to a destination state relative to the state of the vehicle at the time the action to be performed is decided, $t_{0}=0$. We define the relative change associated with each action as a $(dx,dv) \in \mathbb{R}^{2}$ pair of intended changes in x position and velocity, such that:

\begin{equation*}
    \vec{x}_{\text{final}} := 
    \begin{bmatrix}
        x_{\text{final}}\\y_{\text{final}}\\v_{\text{final}}\\\theta_{\text{final}}
    \end{bmatrix}
 = 
\begin{bmatrix}
    x_{t_{0}}\\y_{t_{0}}\\v_{t_{0}}\\\theta_{t_{0}}
\end{bmatrix}
+
\begin{bmatrix}
    dx\\0\\dv\\0
\end{bmatrix}
\end{equation*}

Following the approach specified in \cite{during2014} we represent the trajectory defined by $(\vec{x}_{\text{init}},\vec{u}_{\text{init}},\vec{x}_{\text{final}})$ with a pair of polynomials

\begin{equation*}
    \begin{split}
    x(t) &= x_{5}t^{5}+x_{4}t^{4}+x_{3}t^{3}+x_{2}t^{2}+x_{1}t^{1}+x_{0}\\
    y(t) &= y_{4}t^{4}+y_{3}t^{3}+y_{2}t^{2}+y_{1}t^{1}+y_{0}
    \end{split}
\end{equation*}

such that


\begin{equation*}
    \begin{split}
        x(t_{\text{init}}) &= x_{\text{init}},\quad y(t_{\text{init}}) = y_{\text{init}}\\
        \dot{x}(t_{\text{init}}) &= v_{\text{init}}*cos(\theta_{\text{init}}), \quad \dot{y}(t_{\text{init}}) = v_{\text{init}}*sin(\theta_{\text{init}})\\
        \ddot{x}(t_{\text{init}}) &= a_{x}, \quad \ddot{y}(t_{\text{init}}) = a_{y}\\
        x(T) &= x_{\text{final}} \\
        \dot{x}(T) &= v_{\text{final}}*cos(\theta_{\text{final}}), \quad \dot{y}(T) = v_{\text{final}}*sin(\theta_{\text{final}})\\
        \ddot{x}(T) &= 0, \quad \ddot{y}(T) = 0
    \end{split}
\end{equation*}

where $T$ is the length of the trajectory, $t_{\text{init}}$ is the time when planning is taking place, and $(a_{x},a_{y})$ is the parametric acceleration at time $t_{\text{init}}$. We explicitly presume that upon reaching its destination, the car will have $0$ lateral and longitudinal acceleration. Implicitly, based on the definition for $\vec{x}_{\text{final}}$ we also presume that the vehicle's final heading is the same as the heading at time of planning $t_{0}$. We do not define the $y$ trajectory on the final position, as the change in x-coordinate and velocity are sufficient for the purposes of defining lane changing or lane keeping trajectories.\  

Producing trajectories using this method is computationally tractable allowing for more frequent re-computation, and also facilitates imposing speed limit and jerk constraints on the generated trajectories. The estimated state at any time $0\leq t \leq T$, $\vec{x}^{-i}_{t} := (x_{t},y_{y},v_{t},\theta_{t})^{T}$ along the trajectory can be computed from the parametric state $(x(t),y(t),\dot{x}(t),\dot{y}(t),\ddot{x}(t),\ddot{y}(t))^{T}$. Therefore we can construct a discretisation of the trajectory $X^{-i} = \{\vec{x}^{-i}_{k \Delta t}\}_{k=0}^{k=\frac{T}{\Delta t}}$

\subsection{Trajectory Planning}
We formulate the trajectory planning problem as an optimal control problem where the planning agent minimises an egoistic cost function, conditioned on the discretised estimated trajectory for the other vehicle $X$.

\begin{equation*}
    \begin{aligned}
         \underset{\vec{x}_{0}, \ldots ,\vec{x}_{N+1},\vec{u}_{0},\ldots,\vec{u}_{N}}{\text{minimize}}
        &\sum_{j=0}^{N}J(\vec{x}_{j},\vec{u}_{j},\vec{x}_{\text{final}}^{i}) \\
         \text{subject to}\\
        & \vec{x}_{0} = \vec{x}_{\text{init}}^{i}\\
        & \vec{x}_{k+1} = F(\vec{x}_{k},\vec{u}_{k}) \; k = 0, \ldots, N \\
        & \vec{b}(\vec{x}_{j},\vec{u}_{j}) \leq 0, \; j = 0, \ldots, N.\\
        & \vec{c}(\vec{x}_{j},\vec{x}_{j}^{-i}) \leq 0, \; j = 0, \ldots, N.
    \end{aligned}
\end{equation*}

where $F(\vec{x},\vec{u})$ is the kinematic bicycle model, $\vec{b}$ defines the egoistic constraints on the vehicle, and $\vec{c}$ defines the collision avoidance constraints.\\

We let $J(\vec{x},\vec{u},\vec{x}_{\text{dest}}) = \vec{w}^{T} \cdot (\vec{x}-\vec{x}_{\text{dest}})^{2}$, the sum of squared distance to the intended destination state, where $\vec{w} \in \mathbb{R}^{4}$ are manually specified weights. This cost function drives the planner to the desired destination as quickly as permitted by the constraints, which include collision avoidance constraints to account for the presence of the other vehicle. \ 

In $\vec{b}$ we apply bounds on the x-coordinate to keep the car on the road, speed limit constraints, as well as heading constraints to prevent the car from doing a u-turn on the road. We also bound the control inputs, $(a,\delta)$, to be within reasonable bounds. We adapt the collision avoidance condition from \cite{during2014} and use it as an collision avoidance constraint in $\vec{c}(\vec{x}^{i},\vec{x}^{-i})$;

\begin{equation*}
    1- \left ((\frac{x^{i}_{j} - x^{-i}_{j}}{r_{x}})^{2} + (\frac{y^{i}_{j} - y^{-i}_{j}}{r_{y}})^{2} \right ) < 0
\end{equation*}

where $(r_{x},r_{y}) \in \mathbb{R}^{2}$ are the half-length of the major and minor axes of an ellipse centered on the vehicle.

\section{Experimental Method}
\label{experimental_method}
We run our experiments on a scenario as depicted in Figure \ref{intro_figure}, with car $C1$ in the left lane wanting to merge into the right lane to avoid having to brake, and car $C2$ preferring to continue at their current pace over giving way. In terms of $(dx,dv)$, we define these options as $\{((lw,0),0,-10m/s)\}$ and $\{(0,0),(0,-5m/s)\}$ respectively, where $lw$ is the lane width in metres. For simplicity we do not include the stationary obstacle in the experiment, but the values in the reward matrix are manually constructed, and are independent of the environment, so this omission does not affect planner behaviour. We also changed the values in the reward matrix to be:
\vspace{5pt}

\begin{tabular}{cc|c|c|}
    & \multicolumn{1}{c}{} & \multicolumn{2}{c}{$C2$}\\
    & \multicolumn{1}{c}{} & \multicolumn{1}{c}{$GW$}  & \multicolumn{1}{c}{$C$} \\\cline{3-4}
    \multirow{2}*{$C1$}  & $D$ & $(-1,-1)$ & $(0,1)$ \\\cline{3-4}
    & $LC$ & $(1,0)$ & $(-1,-1)$ \\\cline{3-4}
\end{tabular}
          
\vspace{8pt}
  
\ This does not affect the performance of any of the decision-makers we evaluate since, as can be seen in Section \ref{altruism_and_area_of_conflict}, the reasoning applied only depends on the values in the preferred outcomes for each player, in this case the values on the antidiagonal. This does affect the true reward received by the players, which will be part of our evaluation.\

The initial positions of the cars are randomly perturbed such that either car can start up to a car length ahead of the other, and offset from the middle of the lane by up to a quarter of the lane width. Both cars start with an initial velocity of $15m/s$, with a speed limit of $15m/s$. The acceleration range is $[-3m/s^{2},3m/s^{2}]$ and the rate of change of heading, $\delta$ is bound in $[-1\text{deg}/s^{2},1 \text{deg}/s^{2}]$.\ 

In our experiments we evaluate the performance of interactive decision-makers utilising; SVO~\citep{Schwarting2019}, Altruism and Augmented Altruism. We use a non-interactive decision-maker as a baseline to compare the performance against. To make the results comparable use a finite set of coefficients, and run identical experiments with each combination $(\alpha_{1},\alpha_{2})$. For Altruism and Augmented Altruism  the coefficient values used are:

\begin{equation*}
    A:= [0,.25,.51,.75,.99]
\end{equation*}

where the .51 value was used to avoid stalemates in decision making (i.e. scenarios where a player would not have a clear preference, and .99 used to avoid violating the constraints on Augmented Altruism. Equivalently for SVO:
\begin{equation*}
    SVO := [0,\frac{\pi}{8},\frac{\pi}{4},\frac{3\pi}{8},1]
\end{equation*}

We solve the optimal control trajectory planning problem using Model Predictive Control in a receding horizon fashion. We use a timestep size, $\Delta t$ of $.2$ seconds, a lookahead horizon of $4$ seconds. We use the optimal planning library CasADi(\cite{Andersson2019}) to solve the Optimal Control problem using an Interior Point Optimizer method, replanning at 1Hz.\  

At the start of each experiment each player independently decides upon an action using the same decision-making method and the assigned coefficients. During planning, each player $i$ uses the state of the other player, $-i$, to construct a coarse approximation of their presumed trajectory, $X^{-i}$. Player $i$ then uses this to generate their own optimal trajectory, which they follow. Once the agent has satisfied this initial objective, if it is not their ``true" objective, as specified by the true reward matrix, the agent then plans an optimal trajectory to satisfy this objective. The experiment concludes when both agents have achieved their true objectives, when $\exists t<T \quad s.t. \quad  \parallel~\vec{x}^{i}_{t} - \vec{x}^{i}_{\text{dest}}~\parallel_{2} < \epsilon$. We record the time,$t_{c}^{i}$, when each agent satisfied their objective for the final time (e.g. if an agent satisfies an objective, but then has to move away from it to avoid collision, $t_{c}^{i}$ is reset). The true reward $r_{i}$ the planner converged to for each agent is also recorded.\ 
\end{appendices}

\end{document}